\documentclass[12pt,psfig]{article}
\usepackage{amssymb}
\usepackage{amsmath}
\usepackage{epsfig}
\makeatletter
\def\@cite#1#2{{[{#1}]\if@tempswa\typeout
{IJCGA warning: optional citation argument
ignored: `#2'} \fi}}

\newcount\@tempcntc
\def\@citex[#1]#2{\if@filesw\immediate\write\@auxout{\string\citation{#2}}\fi
  \@tempcnta\z@\@tempcntb\m@ne\def\@citea{}\@cite{\@for\@citeb:=#2\do
    {\@ifundefined
       {b@\@citeb}{\@citeo\@tempcntb\m@ne\@citea\def\@citea{,}{\bf ?}\@warning
       {Citation `\@citeb' on page \thepage \space undefined}}%
    {\setbox\z@\hbox{\global\@tempcntc0\csname b@\@citeb\endcsname\relax}%
     \ifnum\@tempcntc=\z@ \@citeo\@tempcntb\m@ne
       \@citea\def\@citea{,}\hbox{\csname b@\@citeb\endcsname}%
     \else
      \advance\@tempcntb\@ne
      \ifnum\@tempcntb=\@tempcntc
      \else\advance\@tempcntb\m@ne\@citeo
      \@tempcnta\@tempcntc\@tempcntb\@tempcntc\fi\fi}}\@citeo}{#1}}
\def\@citeo{\ifnum\@tempcnta>\@tempcntb\else\@citea\def\@citea{,}%
  \ifnum\@tempcnta=\@tempcntb\the\@tempcnta\else
   {\advance\@tempcnta\@ne\ifnum\@tempcnta=\@tempcntb \else
\def\@citea{--}\fi
    \advance\@tempcnta\m@ne\the\@tempcnta\@citea\the\@tempcntb}\fi\fi}
\makeatother

\newcommand{\gsim}{\lower.7ex\hbox{$\;\stackrel{\textstyle>}{\sim}\;$}}
\newcommand{\lsim}{\lower.7ex\hbox{$\;\stackrel{\textstyle<}{\sim}\;$}}
\newcommand{\be}{\begin{equation}}
\newcommand{\ee}{\end{equation}}
\newcommand{\bea}{\begin{eqnarray}}
\newcommand{\eea}{\end{eqnarray}}

\usepackage{amssymb}

\def\baselinestretch{1}
\begin{document}
\catcode`@=11
\newtoks\@stequation
\def\subequations{\refstepcounter{equation}%
\edef\@savedequation{\the\c@equation}%
  \@stequation=\expandafter{\theequation}
  \edef\@savedtheequation{\the\@stequation}
  \edef\oldtheequation{\theequation}%
  \setcounter{equation}{0}%
  \def\theequation{\oldtheequation\alph{equation}}}
\def\endsubequations{\setcounter{equation}{\@savedequation}%
  \@stequation=\expandafter{\@savedtheequation}%
  \edef\theequation{\the\@stequation}\global\@ignoretrue

\noindent}
\catcode`@=12
\begin{titlepage}
\title{{\bf Compactifications of conformal gravity}}
\vskip2in
\author{
{\bf Ignacio Navarro$$\footnote{\baselineskip=16pt E-mail: {\tt
ignacio.navarro@durham.ac.uk}}} $\;\;$and$\;\;$ {\bf Karel Van
Acoleyen$$\footnote{\baselineskip=16pt E-mail: {\tt
karel.van-acoleyen@durham.ac.uk}}}
\hspace{3cm}\\
 $$ {\small IPPP, University of Durham, DH1 3LE Durham, UK}.
}

\date{}
\maketitle
\def\baselinestretch{1.15}
\begin{abstract}
\noindent

We study conformal theories of gravity, $i.e.$ those whose action
is invariant under the local transformation $g_{\mu\nu}
\rightarrow \omega^2 (x) g_{\mu\nu}$. As is well known, in order
to obtain Einstein gravity in 4D it is necessary to introduce a
scalar compensator with a VEV that spontaneously breaks the
conformal invariance and generates the Planck mass. We show that
the compactification of extra dimensions in a higher dimensional
conformal theory of gravity also yields Einstein gravity in lower
dimensions, without the need to introduce the scalar compensator.
It is the field associated with the size of the extra dimensions
(the radion) who takes the role of the scalar compensator in 4D.
The radion has in this case no physical excitations since they are
gauged away in the Einstein frame for the metric. In these models
the stabilization of the size of the extra dimensions is therefore
automatic.

\end{abstract}

\thispagestyle{empty} \vspace{5cm}  \leftline{}

\vskip-20.5cm \rightline{} \rightline{IPPP/05/08}
\rightline{DCPT/05/16} \vskip3in

\end{titlepage}
\setcounter{footnote}{0} \setcounter{page}{1}
\newpage
\baselineskip=20pt





\section{Introduction: conformal invariance in 4D}

Symmetries play a central role in physics and amongst them local
symmetries are especially important: they reflect the redundancies
that the introduction of coordinates in spacetime or field space
inevitably produce. Local symmetries form the skeleton of our
description of particle interactions since a quantum field theory
involving spin one fields needs gauge symmetry for consistency. In
the same fashion, invariance of the action under diffeomorphisms
is the basis of our understanding of gravity.

However symmetries are often spontaneously broken, or hidden, in
our universe. This happens when the vacuum (or some ``parameters''
of our low energy Lagrangian) transform under the symmetry in
question. This phenomenon occurs with the electroweak interactions
of the standard model, where the gauge group $SU(2)\times U(1)$ is
spontaneously broken down to the electromagnetic $U(1)$, and how
and why it takes place are probably the most prominent questions
in elementary particle theory. In this sense one could also say
that diffeomorphism invariance is spontaneously broken down to
Poincar\'{e} invariance in Minkowski spacetime, since one can see
the metric as a field with a vacuum expectation value (VEV),
breaking the diffeomorphism invariance down to the isometry group
of spacetime.

An interesting question is then if there are other non-apparent
local symmetries under which the spacetime metric transforms. In
this letter we consider one such possibility: conformal theories
of gravity, $i.e.$ those whose action is left invariant under the
transformation \be g_{\mu \nu} \rightarrow \omega (x)^2 g_{\mu
\nu} \label{conf} \ee for any smooth non-zero $\omega(x)$. Notice
that this is not a coordinate transformation: although some
coordinate transformations (dilatations) can have a similar effect
on the metric, this transformation is not related to a change of
coordinates, it is a genuine new local internal symmetry of the
metric. At the quantum level this symmetry, like gauge symmetries,
is anomalous. We will assume in this letter that this anomaly can
be cancelled with the addition of suitably coupled matter fields,
as happens with the gauge symmetries of the Standard Model (SM),
so a quantum theory respecting this symmetry can be built (see for
instance \cite{Fradkin:1985am}). In fact, conformal invariance is
regarded as a property of the theory making the gravitational
quantum corrections more tractable and even as a necessity if one
is to find a ultraviolet renormalization group fixed
point \cite{Smolin:1979uz}. Remember that since the metric produces
our local units of measure, this symmetry seems to be in conflict
with any dimensionful parameter of the Lagrangian, so dimensionful
parameters should conceal fields that transform under this
rescaling.

It is time to recall that we could not write down a realistic
Lagrangian for describing the world without at least two
dimensionful parameters: the scale of electroweak symmetry
breaking ($M_{ew}\sim 10^2 GeV$), given by the Higgs mass in the
SM and the Planck mass ($M_p \sim 10^{18}GeV$), controlling the
strength of gravity. To these we could add the vacuum energy
scale, apparently of the order of $\Lambda_{vac} \sim
(10^{-3}eV)^4$ \cite{Riess:2004nr}. The hugely different
magnitudes of these scales have been disturbing theoretical
physicists for years, and no fully convincing explanation for
these enormous hierarchies has been put forward so far. The
dimensionful parameters of the Lagrangian describing the universe
are thus the most mysterious ones, especially if we consider the
quantum corrections to the theory that naively would seem to
contribute to all of them with large and similar amounts. In this
context, the conformal symmetry could have interesting
implications\footnote{See \cite{Wetterich:1987fm} for a discussion
of cosmology in the context of global conformal invariance.},
since it links all mass scales to a common origin, the breaking of
conformal invariance\footnote{In a sense this breaking is
inevitable, like that of the diffeomorphism invariance. For
finding a theory with a conformally invariant phase one should go
to theories in which the metric is a derived quantity, like
metric-affine gauge theories of gravity \cite{Hehl:1994ue}.}.
Notice also the relation $M_{ew}^2 \sim M_p \Lambda^{1/4}$ that
could be interpreted as pointing to a common origin of these mass
scales.

If we want to build a conformally invariant Lagrangian for gravity
in 4D that involves only the metric, the only option we have is
Weyl gravity \be S_{gravity} \propto \int \sqrt{g}
C_{\mu\nu\lambda\rho}C^{\mu\nu\lambda\rho} d^4 x \ee where
$C^{\mu}_{\;\nu\lambda\rho}$ is the Weyl tensor, that is invariant
under the transformation (\ref{conf}). This theory was proposed by
Weyl \cite{Weyl:1918ib} back in the early days of general
relativity. In a series of papers Mannheim and Kazanas
\cite{Mannheim:1988dj}, have put forward Weyl gravity as an
alternative to Einstein gravity that could also explain the galaxy
rotation curves without the need for dark matter, but such a
theory leaves many unanswered questions and it is not clear that
it can provide a realistic alternative to General Relativity
\cite{Pireaux:2004xb}. Furthermore, even if we give the Higgs
field and the fermion fields weights under the conformal
transformation: \be h\rightarrow
\omega(x)^{-1}h\,,\,\,\,\,\,\,\,\,\,\psi\rightarrow
\omega(x)^{-3/2}\psi\,,\label{confSM} \ee the action of the SM is
not conformally invariant. Specifically, the only symmetry
violating terms are the Higgs kinetic term and the Higgs mass
term. In the fermion kinetic term, the transformation of the spin
connection compensates the derivative terms originating from the
transformation of the fermion fields.

If we want to construct a Lagrangian invariant under the conformal
transformation (\ref{conf}) and still recover Einstein gravity we
must introduce in the theory a compensator field, transforming
under this symmetry like \be \phi \rightarrow \omega (x)^{-1}
\phi. \label{confphi} \ee In this case we can write down an action
for gravity\cite{Dirac:1973gk}, invariant under (\ref{conf}) and (\ref{confphi}), as\footnote{Our conventions are diag$(g_{\mu\nu})=(-+++)$,
$R^{\rho}_{\;\;\mu\lambda\nu}=\partial_{\lambda}\Gamma^{\rho}_{\nu\mu}+\ldots$,
$R=R^{\rho\mu}_{\;\;\;\;\rho\mu}$.}
 \be S_{gravity} = \int
{\rm d}^4 x \sqrt{g}\left\{ \phi^2 R + 6
\partial_\mu \phi
\partial^\mu \phi \right\}.\label{conformalGR}
\ee We can now also use the field $\phi$ to construct a conformally
invariant SM Lagrangian as
\bea S_{SM} = -\int {\rm d}^4
x\sqrt{g}\{\phi^2 {\cal D}_\mu h {\cal D}^\mu h^{\dagger} + \phi^4
V(h)
+\frac{1}{4}F_{\mu\nu}^{(a)} F^{(a)\mu\nu} \nonumber\\
+\phi^{3/2}i{\overline\psi}^i\gamma^\mu(x)(\mathcal{D}_{\mu}+\Gamma_{\mu}(x))(\phi^{3/2}\psi^i)
+\lambda_i\phi^4\overline{\psi}^i_L.h\psi^i_R+h.c.\}, \eea where
the Higgs field, the gauge fields $A_{\mu}^a$ and the fermion
fields are left invariant under the conformal transformation. One
can bring the Lagrangian to a more conventional form with the
redefinitions $h\rightarrow h\phi^{-1}, \psi^i\rightarrow
\psi^i\phi^{-3/2}$. This redefinition 'covariantizes' the
derivatives in the Higgs kinetic term: \be \phi {\cal D}_\mu h
\rightarrow \left({\cal D}_\mu-\frac{\partial_\mu
\phi}{\phi}\right) h\,, \ee and reduces all the other terms to the
conventional ones of the SM except for a coupling of the scalar
compensator to the Higgs mass term. The Higgs field and the
fermion fields now transform according to (\ref{confSM}).

Notice that the kinetic term of the field $\phi$ is ghostlike.
This does not pose any problems since the field actually is a
ghost: its excitations can be gauged away by conformal
transformations. Assuming $\phi$ is not zero, we can use our
conformal gauge freedom to go to a gauge in which $\phi$ is
constant, (applying the conformal transformation with $\omega =
\phi/M_p$) and we recover in this conformal ``unitary'' gauge the
equations of the SM coupled to Einstein gravity. So, after the
assumption of a non-zero VEV for $\phi$, this realization of
conformal invariance appears to be devoid of physical meaning at
the classical level since it just corresponds to taking the
conventional action of the SM and GR, substitute in it $g_{\mu
\nu} \rightarrow \phi^{2} g_{\mu \nu}$ and take independent
variations with respect to $g_{\mu \nu}$ and $\phi$. The equation
of motion of $\phi$ is not independent of the rest, since it is
just the trace of the Einstein equations. We have introduced a new
degree of freedom (and a new symmetry) only to subtract it using
our conformal gauge freedom.

There is however a different realization of conformal invariance
in 4D that actually adds some new degrees of freedom to the
gravity sector, and makes use of a vector field, besides the
scalar compensator, to gauge the conformal
invariance\footnote{This vector field was first introduced by Weyl
\cite{Weyl:1918ib} who tried to identify it with the
electromagnetic potential, and it is sometimes called the Weylon.}
(see also \cite{Iorio:1996ad} for a discussion of a gauging of
conformal invariance that makes use of the transformation
properties of the Ricci tensor). This new vector field transforms
under the conformal symmetry like \be W_\mu \rightarrow W_\mu -
\partial_\mu log \;\omega \ee and has a minimal coupling to
scalars (covariantizing its kinetic terms with respect to
conformal transformations) but does not couple to fermions (see
for instance \cite{Hochberg:1990xp}) and it gets a mass of the
order of $M_p$ as a consequence of the VEV of the scalar
compensator\footnote{It has been suggested that in this case,
economically, the Higgs could be the field taking the role of the
conformal compensator \cite{Cheng:1988zx}, in which case there
would be no physical Higgs in the spectrum. All its degrees of
freedom would be ``eaten'' in the electroweak and conformal
symmetry breaking processes by the massive vector bosons.}.

So we have seen that in 4D there are ways of implementing a
symmetry that conformally transforms the metric (in a realistic
model) at the cost of introducing a scalar that also transforms
under this symmetry and assuming a VEV for it. This expectation
value is the mass scale against which we can make dimensional
measurements in our theory. The assumption of a VEV for this field
is crucial for obtaining a realistic model, but this might be
regarded as an $ad$ $hoc$ way of realizing the conformal symmetry
since it is not clear how to obtain the VEV for this field as the
result of a minimization process. So one might wonder: is it
possible to construct a realistic theory invariant under a
conformal transformation of the metric only? (In a theory in which
other fields, besides the metric, have non-zero weight under
conformal transformations, a positive answer to this question
would mean that one can construct viable models even if these
other fields are set to zero in the background.) In such a model
all spacetimes related by a transformation like (\ref{conf}), {\it
with other fields constant}, would be describing the same state so
the geometry of the (full) spacetime would not be observable. Only
conformally invariant quantities have physical meaning. This seems
quite counter-intuitive but we will see that realistic models with
this property can be built by assuming the existence of
compactified extra dimensions. The size of the extra dimensions
provides us with a mass scale against which we can make (relative)
dimensionful measurements in the 4D effective theory.

In the next section we consider conformal theories of gravity in
higher dimensions.  We consider as the action for gravity the most
general Lagrangian built out of the metric invariant under the
transformation (\ref{conf}).  We show that these theories
generically admit compactifications of the extra dimensions in a
constant curvature manifold while the non-compact dimensions can
have positive, negative or zero curvature. We provide explicit
examples in six and eight dimensions. Furthermore we show that
Einstein gravity is recovered at low energies, and the conformal
invariance is non-linearly realized in 4D with a compensator field
that has a VEV. This field corresponds to the size of the extra
dimensions (the radion) and its excitations can be ``gauged away''
by choosing the Einstein frame for the metric. There is then no
need to consider a stabilization mechanism for this modulus field
in this kind of compactification.

\section{Higher dimensional conformal gravity}

Let us now consider a general conformal invariant action in an
even number ($D$) of dimensions \be S=\int \!\!d^D{\rm x}
\sqrt{G}\left( {\cal L}_{gravity}+\delta {\cal
L}_{matter}\right)\,. \ee ${\cal L}_{gravity}$, generating the
left hand side of the equations of motion (EOM),  will consist of
a linear combination of all conformally invariant local scalar
densities that can be built out of the metric (see appendix). The
number of independent conformally invariant terms increases
rapidly with the dimension \cite{Fulling:1992vm}: in 6 dimensions
there are 3 such terms, in 8 dimensions there are 12 terms, while
in 10 dimensions we have already 67 conformal invariant terms. (It
stills remain to be seen if they are all independent at the level
of the EOM, $i.e$ if we put the total derivatives to zero
\cite{BoulangerII}.) $\delta {\cal L}_{matter}$, producing the
right hand side of the EOM, in the form of the energy momentum
tensor, will consist of that part of the matter Lagrangian
responsible for the compactification.

We will first look for compactifications into a spacetime
background with a factorizable metric like  \be ds^2 = G_{MN}
d{\rm x}^M d{\rm x}^N = g_{\mu\nu}(x) dx^{\mu}dx^{\nu} +
\gamma_{ij}(z)dz^i dz^j \label{metric} \ee where $g_{\mu \nu}$ is
the 4D Lorentzian metric of a maximally symmetric manifold, with
curvature $R_g$, and $\gamma_{ij}$ is the metric for a compact
euclidean $n$-dimensional maximally symmetric space ($n=D-4$),
with curvature $R_{\gamma}$. For this metric the gravity
Lagrangian will take the form (see appendix) \be {\cal L}_{gravity}= {\cal
L}(R_g,R_\gamma)=\sum_{i=0}^{D/2} c_iR_g^{D/2-i}R_\gamma^i\,. \ee
One can now easily obtain the left hand side of the EOM, by
considering variations $g_{\mu\nu}\rightarrow
g_{\mu\nu}(1+\epsilon)$ or $\gamma_{\mu\nu}\rightarrow
\gamma_{\mu\nu}(1+\epsilon)$, for constant $\epsilon$: \bea
W_{MN}&=&\left(\begin{array}{cc}g_{\mu
\nu}(\frac{R_g}{4}\frac{\partial
{\cal L}}{R_g} - \frac{{\cal L}}{2}) & \\
&  \gamma_{ij}(\frac{R_\gamma}{n}\frac{\partial {\cal
L}}{R_\gamma} - \frac{{\cal L}}{2}) \end{array} \right)\label{LEOM}\\
&=&\left(\begin{array}{cc}g_{\mu \nu}(\sum_{i=0}^{D/2}
\frac{c_i}{4}(\frac{D}{2}-i-2)R_g^{D/2-i}R_\gamma^i) & \\
&  -\gamma_{ij}\frac{4}{n}(\sum_{i=0}^{D/2}
\frac{c_i}{4}(\frac{D}{2}-i-2)R_g^{D/2-i}R_\gamma^i)
\end{array} \right).\nonumber
\eea Notice that, because of the conformal symmetry, $W_N^N=0$.
The energy momentum tensor will also be traceless, for the same
reason. We will show later that, with a proper choice for $\delta
{\cal L}_{matter}$, one indeed produces a energy momentum tensor
of the form: \be
T_{MN}=\left(\begin{array}{cc}-g_{\mu \nu}\Lambda & \\
&  \gamma_{ij}\Lambda\frac{4}{n} \end{array} \right).\label{emt}
\ee So, as a consequence of the conformal invariance, the two EOM
one has in the case of conventional factorizable compactifications (of Einstein
gravity), now boil down to one. For a given $\Lambda$, the
curvatures $R_\gamma$ and $R_g$ are not uniquely determined. It is
the (conformally invariant) ratio of $\Lambda$ and
$R_\gamma^{D/2}$ that determines the curvature of the 4D world. In
fact it does not make sense to talk of a value for $\Lambda$ until
we have fixed the conformal gauge, since under a conformal
transformation this parameter changes as $\Lambda \rightarrow
\omega^{-D} \Lambda$. We can fix the conformal gauge by requiring
$R_\gamma$ to be an arbitrary constant, and it is then apparent
that only for a particular value of $\Lambda$ in this gauge,
namely $\Lambda=\frac{c_{D/2}}{2}R_\gamma^{D/2}$, we recover flat
4D space.

We will now consider some examples of conformally invariant matter
Lagrangians in 6 or 8 dimensions that produce an energy momentum
tensor as in (\ref{emt}). Just as in Einstein gravity we will use
n-forms \cite{Freund:1980xh} or scalars \cite{Gell-Mann:1984sj} to
produce the spontaneous compactification, but we will have to add
them in such a way that the conformal invariance is preserved. Any
matter field will be considered invariant under the conformal
symmetry, holding the promise of keeping the transformation
(\ref{conf}) as a purely geometric space-time transformation.

In $D$ dimensions a natural field to consider is an antisymmetric
$(\frac{D}{2}-1)$-form ($\bf{A}$) whose action \be \delta
S_{matter} = -\frac{1}{D/2!}\int{\rm d}^D {\rm x} \sqrt{G}
H^{M..N}H_{M..N}\,, \ee with ${\bf H} = d\bf{A}$, is invariant
under (\ref{conf}). So in 8D we can just consider a 3-form field
$A_{MNP}$ and consider a vacuum expectation value of its field
strength as
\begin{eqnarray}
H_{\mu\nu\lambda\sigma}=\sqrt{g}E \epsilon_{\mu\nu\lambda\sigma}\nonumber \\
H_{ijkl}=\sqrt{\gamma}B \epsilon_{ijkl}
\label{fluxes}\end{eqnarray} with $\epsilon_{abcd}$ the
Levi-Civita symbol with 4 indices, $E$ and $B$ constants and the
rest of the components of ${\bf H}$ zero. This ansatze solves the
equation of motion for the form, and yields an energy-momentum
tensor given by \be
T_{MN}=\left(\begin{array}{cc} -g_{\mu \nu}\frac{E^2 + B^2}{2} & \\
&  \gamma_{ij}\frac{E^2 + B^2}{2} \end{array}
\right).\label{emtF4} \ee If we now take for instance
$\mathcal{L}_{gravity}=\alpha(C_{ABCD}C^{ABCD})^2$, we can use the
value of this Lagrangian in the background (\ref{metric}) (see
appendix) to obtain the EOM: \be
\frac{\alpha}{441}(R_g+R_\gamma)^3(R_\gamma-R_g)=E^2+B^2\,.\label{tune8d}
\ee This gives a flat 4D space if we tune the fluxes to \be
E^2+B^2=\frac{\alpha}{441}R_\gamma^4.\ee Notice that for this
value of the fluxes, one also has the real solution
$R_g=R_g^\star=0.84R_\gamma$. It is tempting to speculate on the
construction of a inflationary scenario of the type discussed in
\cite{Starobinsky:1980te} that might arise from dynamical
solutions starting in $R_g\approx R_g^\star$ and ending up in
$R_g\approx 0$.

In the 6D case, the corresponding conformally invariant 2-form
$A_{MN}$ would naturally produce a (3+3)-splitting of the
energy-momentum tensor, as opposed to the (4+2) we would like to
find in order to generate the compactification to 4D. We should then consider more complicated interactions
in order to yield the required energy-momentum tensor. As an
example we will consider here three possibilities that involve the
addition of a $U(1)$ vector field ($A_M$) and a (spherical or
hyperbolic) sigma model with two scalar fields ($\Phi^i$). The
conformally invariant action is\footnote{Notice that conventional kinetic terms for $\Phi^i$ or $A_M$ are forbidden by the conformal invariance.}
\begin{eqnarray}
\delta S_{matter} = -\int d^{\;6} x\, \sqrt{G} & \{
\frac{\lambda}{4}\left(f_{ij}(\Phi)\partial_M \Phi^i
\partial^M \Phi^j\right)^3+\frac{\kappa}{2} f_{ij}(\Phi)\partial_M \Phi^i
\partial^M \Phi^j F_{NP}F^{NP} \nonumber\\
& + \beta \, C^{MNPQ}F_{MN}F_{PQ}\},
\end{eqnarray}
with $F_{MN} = \partial_{[M}A_{N]}$ and $f_{ij}$ is the metric of
a manifold, with constant curvature $s=\pm1$, parameterized by the
scalar fields. For $\gamma_{ij}(x)$ proportional to $f_{ij}(x)$,
it can be checked that a solution for the scalar fields and the
$U(1)$ field is simply \be \Phi^{i} = z^i \;\;\; ; \;\;\; F_{ij} =
\sqrt{\gamma} B \epsilon_{ij}. \label{scalflux}\ee These VEVs
produce a traceless energy momentum tensor like (\ref{emt}) with
$n=2$ where \be \Lambda=\lambda \left|R_\gamma\right|^3+\kappa B^2
|R_\gamma| + \beta
B^2\left(\frac{R_g}{20}+\frac{3R_\gamma}{5}\right)\label{emt6D}
\,.\ee If we now take for instance $\mathcal{L}_{gravity}=\alpha
C^{A \;\; C}_{\; \; B \;\; D}C^{B \;\; D}_{\; \; E \;\; F}C^{E
\;\; F}_{\; \; A \;\; C}$, the EOM reads \be
\frac{17\alpha}{57600}(R_g+6R_\gamma)^2(3R_\gamma-\frac{R_g}{4})=\Lambda\,.
\ee We can recover a flat 4D space by tuning the flux to: \be
B^2=\frac{\frac{51\alpha}{1600}-s\lambda}{\frac{3}{5}\beta+s\kappa}R_\gamma^2\,.\label{tune6d}\ee
For certain values of the couplings, this flux will again produce
additional real solutions with non-zero $R_g$.

It is also easy to see that at low energies gravity will be
described by Einstein gravity with conformal invariance
non-linearly realized in the action like in
eq.(\ref{conformalGR}), and the radion field will play the role of
the compensator field. For seeing this we can consider the metric
ansatze \be ds^2 = g_{\mu\nu}(x) dx^{\mu}dx^{\nu}
+\phi(x)^{-2}\gamma_{ij}(z)dz^i dz^j \ee with arbitrary
$g_{\mu\nu}(x)$ and $\phi(x)$ (while $\gamma_{ij}$ would still
correspond to a compact constant curvature manifold, with
curvature $s=\pm1$) and compute the effective 4D action
$S_{(4D)}(g_{\mu\nu},\phi)$. This action will be invariant under
the transformations $g_{\mu\nu}\rightarrow \omega(x)^2 g_{\mu\nu}$
and $\phi \rightarrow \omega(x)^{-1}\phi$ as a consequence of the
higher dimensional conformal invariance, so in order to find
$S_{(4D)}(g_{\mu\nu},\phi)$ we can first compute
$S_{(4D)}(g_{\mu\nu},\phi_0)$, with $\phi_0$ a constant. With a
suitable parametrization of the fluxes ($B=b\times
\phi_0^{dim(B)}$), this will result in an expression of the form
\be S_{(4D)}(g_{\mu\nu},\phi_0)= \int {\rm d}^4x\sqrt{g} \left(a_0
\phi_0^{4}+a_1\phi_0^{2}R+(a_2R^2+
a_3R_{\mu\nu\rho\sigma}R^{\mu\nu\rho\sigma}+a_4R_{\mu\nu}R^{\mu\nu})+\ldots)\right)\,.\label{S4D}\ee
We can now substitute $g_{\mu\nu}\rightarrow
(\phi/\phi_0)^{2}g_{\mu\nu}$ to recover
$S_{(4D)}(g_{\mu\nu},\phi)$: \be S_{(4D)}(g_{\mu\nu},\phi)= \int
{\rm d}^4x\sqrt{g} \left(a_0 \phi^{4}+a_1(\phi^2 R + 6
\partial_\mu \phi
\partial^\mu \phi)+ \ldots \right)\,.\ee
However, this exercise is not very useful, since the conformal
invariance allows us to take $\phi = \phi_0$ constant without loss
of generality. This corresponds to taking the Einstein frame in
the 4D action, and it is clear now why in this frame the
excitations of the volume of the extra dimensional manifold are
the degree of freedom of the higher dimensional metric sacrificed
to fix the conformal gauge. If we tune the fluxes and/or couplings
of $\delta \mathcal{L}_{matter}$ such that the first term in
(\ref{S4D}) disappears, we recover Einstein gravity with a zero
cosmological constant at low energies, since higher order
curvature corrections have a negligible impact on low energy
physics in flat space.

As an example we take the 8D action that we considered before \be
S = \int{\rm d}^8x\sqrt{G}\left\{ \alpha
\left(C_{ABCD}C^{ABCD}\right)^2-\frac{1}{4!}H_{ABCD}H^{ABCD}\right\}
 \,.\ee In the flux background (\ref{fluxes}), with a spherical compactification, this yields an effective 4D action\footnote{This effective action is not exactly the one obtained from the higher dimensional action with the ansatze substituted in it. There is a sign flip in the term with the 'electric' four-form flux, see also the comments in \cite{Navarro:2004mm}.}
\be S_{(4D)} =  V_{\gamma} \int{\rm d}^4x \sqrt{g}\left\{ \alpha
\left(\frac{\phi_0^4}{21}+\frac{2}{21}\phi_0^2R+\frac{R^2}{21}+R_{\mu\nu\rho\sigma}R^{\mu\nu\rho\sigma}-\frac{2}{3}R_{\mu\nu}R^{\mu\nu}\right)^2
-(E^2+B^2)\right\}\,. \ee If we now plug in the volume of the
4-dimensional sphere
$V_{\gamma}=\frac{8\pi^2}{3}(\frac{12}{\phi_0^2})^2$ and tune the
fluxes to the value obtained in (\ref{tune8d}), we indeed recover
Einstein gravity:\be S_{(4D)} =\int{\rm d}^4x
\sqrt{g}\left(\frac{M_p^2}{16\pi}R+\ldots\right)\,,\ee with zero
cosmological constant and a Planck mass
$M_p^2=\frac{8192}{147}\alpha\pi^3\phi_0^2=\frac{65536}{49}\alpha\pi^4\sqrt{2/(3V_\gamma)}$,
set by the volume of the extra dimensional manifold.

For the 6D action that we considered before, the spherical
compactification in the background (\ref{scalflux}), with the fine
tuning (\ref{tune6d}), leads in a similar way to Einstein gravity
with zero cosmological constant and \be
M_p^2=128\pi^2\phi_0^2\frac{\alpha\frac{51}{1600}(2\beta+5\kappa)+\lambda\beta}{3\beta+5\kappa}=\frac{2(8\pi)^3}{V_\gamma}\frac{\alpha\frac{51}{1600}(2\beta+5\kappa)+\lambda\beta}{3\beta+5\kappa}\,.\ee
We see that in these models, although one would naturally expect
the size and curvature of the extra dimensions to be of the order
of the Planck mass, one can also have low compactification scales
(if $\; 3\beta \simeq - 5 \kappa$ in the previous example for
instance).

\section{Conclusions}

In this letter we have considered conformally invariant theories
of gravity as possible extensions of General Relativity motivated
by their higher degree of symmetry. The usual mechanism to recover
Einstein gravity as a long distance effective theory, involved the
introduction of a scalar that also transformed under the conformal
symmetry, only to gauge it away by using our conformal gauge freedom. This seemed to spoil the geometric nature of the
conformal invariance. Furthermore, one had to assume a nonzero VEV
for this field, without any real justification.

We provide an alternative mechanism, that starts from a {\em pure}
conformally invariant theory in higher dimensions. 4D Einstein
gravity is now recovered through a spontaneous compactification
induced by an appropriate matter Lagrangian. This gives a
geometric origin to the scalar compensator field: it is the
radion, or the field associated with the size of the extra
dimensions. Once we take the extra compact dimensions for granted,
the nonzero VEV of this field becomes evident. If the extra
dimensions have spherical topology for instance, the curvature of
the extra dimensions can be arbitrarily small but never zero, so
zero values for the radion are excluded. One can now safely use
the conformal symmetry to fix the value of the compensator field
to a constant. From the higher dimensional point of view, this
reveals the impossibility of destabilization of the size of the
extra dimensions, since we can always use our conformal invariance
to go to a gauge in which the extra dimensions have a fixed size
and curvature. It is precisely in this gauge (the so-called
Einstein gauge) that we recover the canonically normalized
Einstein-Hilbert Lagrangian as the effective action for the metric
in 4D, with the Planck mass proportional to the curvature of the
extra-dimensional manifold. But although the curvature (or the
size) of the extra dimensions can be considered to take a fixed
value without loss of generality, there could be other sources of
instability in these compactifications. The requirement of
stability could restrict the available parameter space in these
models. A deeper study of these issues is however beyond the scope
of the present paper.

\section*{Acknowledgements}K.V.A. was supported by the Fund For
Scientific Research Flanders (Belgium).

\section*{Appendix }
In six dimensions the three conformally invariant local
scalar densities that can be built out of the metric are \cite{Deser:1993yx}:
\begin{eqnarray}
\mathcal{L}^{(6)}_1 &=& C^{A \;\; C}_{\; \; B \;\; D}C^{B \;\;
D}_{\; \; E \;\; F}C^{E \;\; F}_{\; \; A \;\; C}\,,
\nonumber\\ \mathcal{L}^{(6)}_2 &=& C^{A \;\; C}_{\; \; B \;\; D}C^{B \;\; E}_{\; \; A \;\; F}C^{D \;\; F}_{\; \; C \;\; E} \,,\nonumber \\
\mathcal{L}^{(6)}_3 &=& C^{A \;\; C}_{\;\; B \;\; D} \Box C^{B
\;\; D}_{\;\; A \;\; C} + 2 C^{ABCD}C_{ABCE}R^{E}_{D} -
3C_{ABCD}R^{AC}R^{BD}\nonumber\\&& -
\frac{3}{2}R^{AB}R_{BC}R^{C}_A +
\frac{27}{20}R^{AB}R_{AB}R-\frac{21}{100}R^3\,\,.\end{eqnarray} In
the spacetime background (\ref{metric}) they reduce to: \bea
\frac{57600}{17}\mathcal{L}^{(6)}_1=-\frac{14400}{13}\mathcal{L}^{(6)}_2=(R_g+6R_\gamma)^3\,,\nonumber\\
\mathcal{L}^{(6)}_3=\frac{41}{2400}R_g^3-\frac{117}{400}R_g^2R_{\gamma}+\frac{129}{200}R_gR_\gamma^2+\frac{9}{100}R_\gamma^3\,\,.\eea
In eight dimensions there are seven independent Weyl invariants
that do not involve derivatives \cite{Fulling:1992vm,Boulanger:2004zf}:\bea
\mathcal{L}^{(8)}_1&=&(C^{ABCD}C_{ABCD})^2 \,,\nonumber\\
\mathcal{L}^{(8)}_2&=&C^{ABCD}C_{ABC}^{\;\;\;\;\;\;\;E}C^{FGH}_{\;\;\;\;\;\;\;D}C_{FGHE}\,,\nonumber\\
\mathcal{L}^{(8)}_3&=&C^{ABCD}C_{AB}^{\;\;\;\;\;EF}C_{EF}^{\;\;\;\;\;GH}C_{CDGH}\,,\nonumber\\
\mathcal{L}^{(8)}_4&=&C^{ABCD}C_{AB}^{\;\;\;\;\;EF}C_{CE}^{\;\;\;\;\;GH}C_{DFGH}\,,\\
\mathcal{L}^{(8)}_5&=&C^{ABCD}C_{AB}^{\;\;\;\;\;EF}C_{C\;\;E}^{\;\;G\;\;H}C_{DFGH}\,,\nonumber\\
\mathcal{L}^{(8)}_6&=&C^{ABCD}C_{A\;\;C}^{\;\;E\;\;F}C_{E\;\;F}^{\;\;G\;\;H}C_{BGDH}\,,\nonumber\\
\mathcal{L}^{(8)}_7&=&C^{ABCD}C_{A\;\;C}^{\;\;E\;\;F}C_{E\;\;B}^{\;\;G\;\;H}C_{FGDH}\,.\nonumber\eea
Just as for six dimensions, they are all proportional to an
identical term, in the background (\ref{metric}):  \bea
(R_g+R_\gamma)^4&=&441\mathcal{L}^{(8)}_1=3528\mathcal{L}^{(8)}_2=\frac{148176}{13}\mathcal{L}^{(8)}_3=\frac{296352}{13}\mathcal{L}^{(8)}_4\nonumber\\
&=&\frac{592704}{13}\mathcal{L}^{(8)}_5=\frac{1185408}{509}\mathcal{L}^{(8)}_6=\frac{296352}{25}\mathcal{L}^{(8)}_7\,.\eea
The five other Weyl invariants, involving derivatives, were
obtained in \cite{Boulanger:2004zf}.


\end{document}